\documentclass{symp240}

\usepackage{graphicx}
\usepackage{natbib}

\title[Pipeline Reduction of Binary Light Curves]
{Pipeline Reduction of Binary Light Curves from Large--Scale Surveys}

\author[A. Pr\v sa \& T. Zwitter]
{Andrej Pr\v sa$^{1,2}$ \and Toma\v z Zwitter$^2$
\affiliation{$^1$Villanova University, Dept.~of Astronomy, 800 Lancaster Ave, Villanova, PA 19085, USA \\[\affilskip]
$^2$University of Ljubljana, Dept.~of Physics, Jadranska 19, SI-1000 Ljubljana, EU \break 
emails: andrej.prsa@fmf.uni-lj.si, tomaz.zwitter@fmf.uni-lj.si\\
}
}

\pubyear{2007}
\volume{240}  
\pagerange{119--126}
\date{?? and in revised form ??}
\jname{Binary Stars as Critical Tools \& Tests\\ in Contemporary Astrophysics}
\editors{W.I.\ Hartkopf, E.F.\ Guinan \& P.\ Harmanec, eds.}

\begin{document}

\maketitle

\begin{abstract}

One of the most important changes in observational astronomy of the 21$^{\rm st}$ Century is a rapid 
shift from classical object-by-object observations to extensive automatic surveys. As CCD detectors 
are getting better and their prices are getting lower, more and more small and medium-size 
observatories are refocusing their attention to detection of stellar variability through systematic 
sky-scanning missions. This trend is aditionally powered by the success of pioneering surveys such 
as ASAS, DENIS, OGLE, TASS, their space counterpart Hipparcos and others. Such surveys produce massive 
amounts of data and it is not at all clear how these data are to be reduced and analysed. This is 
especially striking in the eclipsing binary (EB) field, where most frequently used tools are 
optimized for object-by-object analysis. A clear need for thorough, reliable and fully automated 
approaches to modeling and analysis of EB data is thus obvious. This task is very difficult because 
of limited data quality, non-uniform phase coverage and solution degeneracy. This paper reviews 
recent advancements in putting together semi-automatic and fully automatic pipelines for EB data 
processing. Automatic procedures have already been used to process Hipparcos data, LMC/SMC observations, 
OGLE and ASAS catalogs etc. We discuss the advantages and shortcomings of these procedures.


\end{abstract}

\firstsection

\section{Introduction}

Doing astronomy today is simply unimaginable without computers. To facilitate observing preparations, 
we use databases; to observe, we use control software; to reduce the acquired data, we use reduction 
programs. Just how far the computer autonomy of the data acquisition process goes is best described 
by the increasing trend of refurbrishing small and medium-size telescopes into fully automatic, 
robotic instruments\footnote{A comprehensive list of more than a hundred such facilities may be 
found, e.g., at {\tt http://www.astro.physik.uni-goettingen.de/\~{}hessman/MONET/links.html}.}. 
Surveys such as OGLE \citep{udalski1997}, EROS \citep{pd1998}, ASAS \citep{pojmanski2002}, space 
mission Hipparcos' epoch photometry \citep{hipparcos1997}, and others, have changed observational 
astronomy: streams of data produced by automatic telescopes around the world and in space are 
overwhelming for currently existing tools and astronomers cannot cope anymore.

Take eclipsing binaries, for example. So far there have been about 500 published papers with physical 
and geometrical parameters determined to better than 3\% accuracy. For a skilled eclipsing binary guru 
it takes 1--2 weeks to reduce and analyse a single eclipsing binary by hand. To date, there are about 
10\,000 photometric/RV data-sets that in principle allow modeling to a 3\% accuracy. By 2020, the 
upcoming missions such as Pan-Starrs \citep{panstarrs2002} and Gaia \citep{perryman2001} will have 
pushed this number to $\sim$10\,000\,000. Even if all observational facilities collapsed at that 
point so that no further data got collected, it would take 12\,500 astronomers to analyse these data 
in the next 100 years! Since there are currently about 13\,000 members of the IAU, the only way to 
achieve this in the next 100 years by traditional methods is to have \emph{every} astronomer in the 
world doing eclipsing binaries. And of course, do not forget to shut down all robotic telescopes out 
there!

With the change in observational astronomy, traditional analysis methods and tools need to change too. 
This paper overviews most important aspects of automatic procedures, tiers that form a pipeline 
reduction of eclipsing binary light curves. Next Section deals with basic principles of the reduction 
and analysis pipeline; Section 3 reviews most important applications of automatic pipelines on 
large-scale survey data. Section 4 stresses the everlasting importance of dedicated observations. 
Finally, Section 5 concludes and gives some prospects for the future.

\section{Tiers of the reduction and analysis pipeline}

\begin{figure}
\begin{center}
\includegraphics[width=6cm]{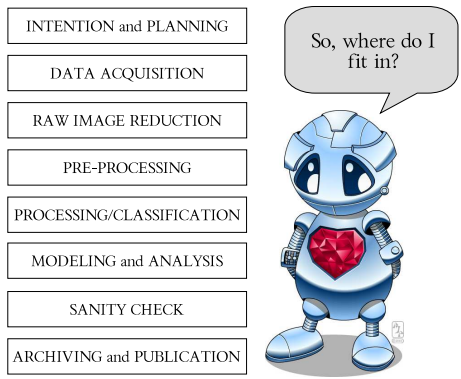} \\
\end{center}
\caption{\label{pipeline} Schematic view of a typical EB reduction and analysis pipeline.}
\end{figure}

A full-fledged pipeline for reduction and analysis of photometric data of eclipsing binary stars 
would ideally consist of 8 distinct tiers depicted in Figure~\ref{pipeline}.

\subsection{Intention and planning}

For as long as we discuss stellar objects in general, and eclipsing binaries in particular, there 
are two apparently frightening facts that need to be considered: {\bf 1)} a target star has already 
been observed and {\bf 2)} a target star has already been observed many times. There are literally 
hundreds of photometric survey missions that have been swiping the sky across and over in a very 
wide magnitude range, and chances are indeed slim that a given star has not been observed yet.

According to Hipparcos results, there are about 0.8\% of eclipsing binaries in the overall stellar 
population (917 out of 118\,218 stars, \citealt{hipparcos1997}). Projecting these statistics to other 
large surveys gives an estimate of how many eclipsing binaries are expected to be present in survey 
databases: $\sim$136\,000 in ASAS ($11\,076$ detected by \citealt{paczynski2006}), $\sim 56\,000$ in 
the OGLE \object{LMC} field ($2\,580$ detected by \citealt{wyrzykowski2003b}), $\sim 16\,000$ in OGLE SMC 
field ($1\,350$ detected by \citealt{wyrzykowski2004}), $\sim$80\,000 in TASS \citep{droege2006} 
etc. Gaia will make a revolution in these numbers since the aimed census of the overall stellar 
population is $\sim$ 1 billion up to $V = 20$ \citep{perryman2001}. Admittedly, magnitude levels and 
variability detection threshold change from survey to survey, but a shortage of eclipsing binaries 
in the databases is more than obvious. In other words, there are many eclipsing binaries out there 
that are either undetected, unconfirmed or misclassified. Stressing a well-known fact that eclipsing 
binaries are unique in their potential to yield accurate masses, radii, temperatures and distances, 
and realizing that many of them are reachable by small-size ground instruments, eclipsing binaries 
should definitely hold one of the top positions on observational candidates list.

\subsection{Data acquisition}

\begin{figure}
\begin{center}
\includegraphics[width=8cm]{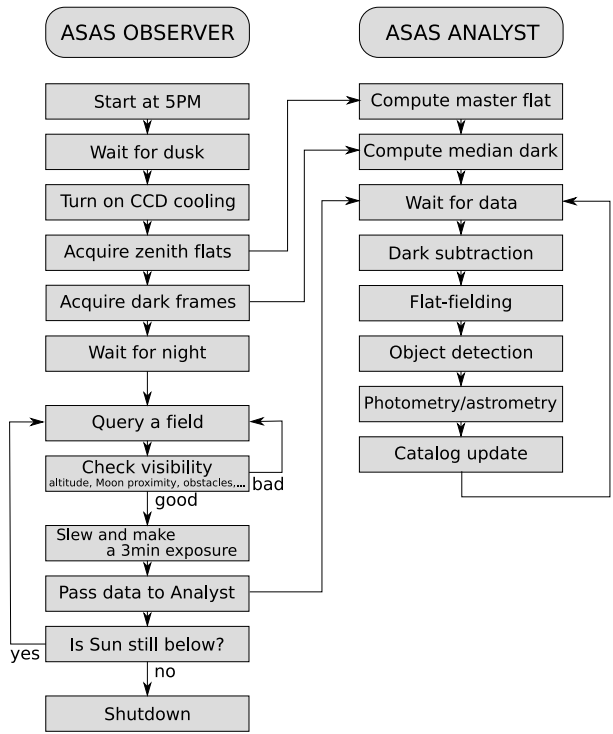} \\
\end{center}
\caption{\label{asas_pipeline} Automatic pipeline of the ASAS project. The pipeline consists of two 
separate (yet connected) engines: Observer and Analyst. The Observer takes care of the data 
acquisition, and the Analyst takes care of data reduction and analysis. The only human intervention 
needed is closing the observatory in case of bad weather and changing the DAT-2 storage tapes. The 
schematic view was adapted from the description of the ASAS project \citep{pojmanski1997}.}
\end{figure}

Most automated of all pipeline tiers, data acquisition has become a truly reliable run-of-the-mill. 
An example of a fully automatic data acquisition and analysis pipeline is that of the All-Sky 
Automated Survey (ASAS, \citealt{pojmanski1997}), depicted on Figure~\ref{asas_pipeline}. The level of 
sophistication is already such that it assures accurate and reliable data both from ground-based and 
space surveys -- and in plenty. A more serious problem for space surveys seems to be telemetry: how 
do we get the data down to Earth? For instance, Hipparcos' downlink rate was only 24 kbit/s, Gaia's 
will be 5 Mbit/s \citep{lammers2005} -- significantly less than the bandwidth we are used to from 
everyday life. To avoid using lossy compression algorithms, surveys must use optimized telemetry for 
the given field and/or data pre-processing (e.g.~binning, filtering, selective downloads). Reliable 
and lossless I/O pipelines and finding ways to store all the acquired data are definitely two of the 
greatest challenges for data acquisition of the future.

\subsection{Raw image reduction}

Acquired data must be reduced: two-dimensional images must be converted to the observed quantity 
(magnitudes, fluxes, \dots). To fully appreciate the need for accurate image reduction, one must 
consider a multitude of physical and instrumental effects that influence the observed data. Some of 
them -- e.g., telescope optics, CCD quantum efficiency and non-linearity, filter response -- may be 
adequately treated during the reduction process. Others -- sky variability, instrumental temperature 
dependence, cosmic rays, interstellar and atmospheric extinction -- usually demand more involvement 
because of their dependence on time and wavelength, or because of unknown physical conditions. Raw 
image reduction consists of taking the acquired image, extracting the data and removing all 
instrumental artifacts contained in that data. This procedure, along with the developed tools (e.g., 
IRAF, \citealt{tody1986}), relies somewhat on human intervention, but in principle it could be 
automated to meet the accuracy of today's surveys. One of the steps in the ASAS pipeline, for example, 
is a fully automated reduction (c.f., Figure~\ref{asas_pipeline}): subtracting dark current and 
flat-fielding \citep{pojmanski1997}.

\subsection{Pre-processing}

Once the images have been reduced, the data are ideally free from instrumental systematics, but 
imprints of other effects (most notably atmospheric extinction and variable seeing) in phased data 
are still present. These effects may be significant and, as such, they should be removed from the 
data. To this procedure we refer to as pre-processing.

There are two approaches to pre-processing: \emph{parametric modeling} and \emph{detrending}. The 
former uses modeling functions and seeks optimal parameters to reproduce the effect at hand; since 
it relies on physical insights, its application is more-or-less transparent. Detrending, on the 
other hand, is based on statistical properties of the observed time series and uses mathematical 
tools to achieve the same goal. Treating atmospheric extinction with parametric models is given 
e.g., by \citet{prsa2005a}, while detrending is presented e.g.~by \citet{tamuz2005}. Since the 
application of the latter is not limited to just a given physical effect, it is well worth stressing 
its major strengths.

\begin{figure}
\begin{center}
\includegraphics[width=\textwidth]{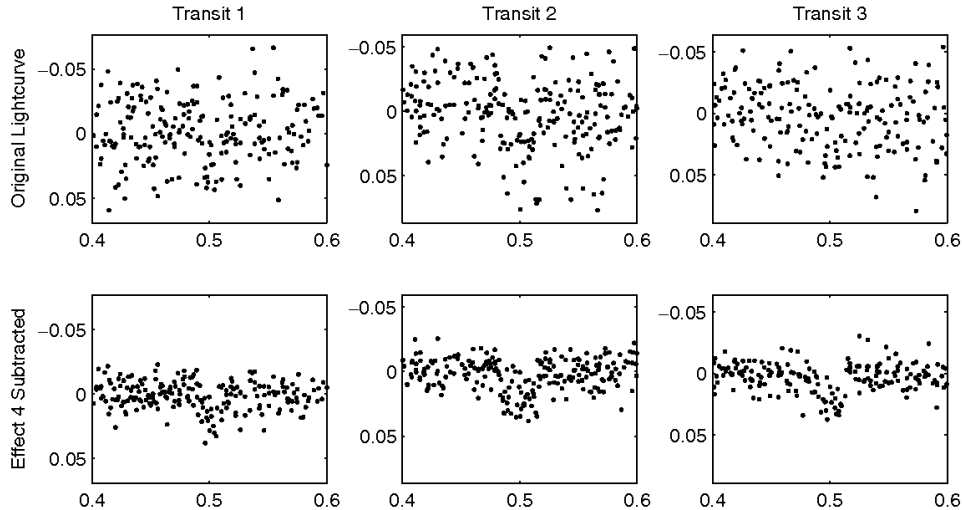} \\
\end{center}
\caption{\label{tmz} Detrending results for 3 planetary transit examples. The plotted diagrams 
depict relative magnitude vs.~phase. The top row shows the original data, and the bottom row shows 
the detrended data, after 4 consecutive detrending iterations. Adapted from \citet{tamuz2005}.}
\end{figure}

Strictly speaking, a \emph{trend} in a time series is a slow, gradual change in observables that 
obscures parameter relationships under investigation. \emph{Detrending} is a statistical operation 
of removing stochastical dependence in consecutive observations, thus making the pre-processed data 
distributed according to the normal (Gaussian) probability distribution function. \citet{tamuz2005} 
proposed a generalized Principal component analysis (PCA) method that accounts for variable 
observation uncertainties. The method is able to remove systematics from the data without any prior 
knowledge of the effect. Figure~\ref{tmz} shows an example of how the algorithm is able to process 
noisy planetary transit data (top row) by consecutively detrending four distinct systematic effects, 
yielding the detrended data (bottom row). Strengths of the method are its universality and little 
importance of the starting values of trend parameters, and the reduction to ordinary PCA in case of 
constant observation uncertainties. Its deficiencies are non-orthogonal eigenvectors (and thus 
deteriorated statistical properties in cases of a highly variable S/N ratio), a danger of filtering 
out intrinsic long-term variability and no relation to physical background of the trend. That said, 
generalized PCA method has proven to be one of the most successful methods for detrending that has 
been applied so far.

\subsection{Processing/classification}

By the time the observed data is ready for scientific munching, most of the non-intrinsic artifacts 
should have been removed. By \emph{processing} we refer to seeking broad scientific properties of 
the observed object: analysis of variance, period determination, phased curve folding etc. Closely 
related is \emph{classification}: based either on the processing results or on statistical pattern 
analysis, the observed objects are classified into their respective groups. While manual approaches 
usually rely on the former principle (we recognize the shape of the light curve and evaluate it 
critically in a broader physical context -- do parameters make sense, is the period plausible for a 
given type of object, \dots), automatic approaches will prefer the latter principle, e.g., through 
the use of Fourier fitting, inversions, neural networks etc. Either way, processing and classification 
aim to discriminate gems from ordinary rocks in terms of our primary interest.

\subsection{Modeling and analysis}

Computationally most demanding task, at least with respect to eclipsing binaries, is their modeling 
and analysis. Seeking and interpreting a set of physical and geometrical parameters involves solving 
the inverse problem. There are many dedicated codes that enable accurate modeling -- WD \citep{wd1971}, 
WINK \citep{wood1971}, NDE \citep{nelson1972}, EBOP \citep{etzel1981}, FOTEL \citep{hadrava1990} and 
many others. We discuss their usage and application to survey data in detail in the following section.

\subsection{Sanity check}

A famous statement by R.E.~Wilson, ``There is more to modeling eclipsing binaries than parameter 
fitting," pretty much encapsulates the idea of sanity check. Solving the inverse problem does not 
only mean finding physical and geometrical parameters that best reproduce the data, it also means 
seeking parameter inter-dependencies, understanding hyperspace non-linearity and, above all, being 
aware of the limitations of the data-set at hand and the used modeling engine. Since eclipsing 
binaries are used for ``calibrating the calibrations", mis- and over-interpreting the data may have 
tragic consequences on solution reliability. Getting a solution from a model is only a fraction of 
the work; the majority is assessing its uniqueness and physical feasibility of that solution.

\subsection{Archiving and publication}

More important than the publication of papers themselves is the question on publishing data. What to 
do with the immense data flow that is expected from large-scale surveys? How to set standards and 
specifications for publishing and storing data? How to coordinate efforts and how to distribute the 
results? Finally, what is our next step in terms of model enhancements? Let us face it -- missions 
such as CoRoT \citep{corot2002} and Kepler \citep{koch2004} will deliver milli-magnitude accuracies 
in just a few years -- do we honestly believe that our models can support such accuracies? All of 
these are still open questions that demand our immediate attention.

\section{First bites on large databases}

One of the first attempts to survey eclipsing binaries in the LMC goes back to \citet{payne1971}, 
who visually examined about 2000 photographic plates, and classified and listed the main 
characteristics of 78 eclipsing binaries. At that time computers only started infiltrating modern 
astronomy and automatic handling was not possible. Yet at the same time, the first EB modeling codes 
were emerging, most notably those of \citet{horak1966,horak1970}, \citet{wd1971}, \citet{wood1971}, 
\cite{nelson1972}, \citet{mochnacki1972} and somewhat later \citet{hill1979}, \citet{etzel1981}, 
\citet{hadrava1990} and \citet{linnell1994}, that would eventually form the base of automatic 
pipelines.

In the early nineties, surveys began to yield first databases that were used for EB detection and 
analysis. \citet{grison1995} assembled a list of 79 EBs in the bar of the LMC from the EROS survey 
data. Of those, only one system was previously identified as an EB, so this work effectively doubled 
the number of known EBs in the LMC. In the year that followed, \citet{friedemann1996} used IRAS data 
\citep{iras1984} to look for coincidences in the positions of EBs taken from the 4$^{\rm th}$ edition 
of the GCVS \citep{gcvs1992} and about 250\,000 IRAS sources. They found 233 candidates, of those 63\% 
Algol-type binaries where accretion disks could be responsible for the IR imprint.

\begin{figure}
\begin{center}
\includegraphics[width=\textwidth]{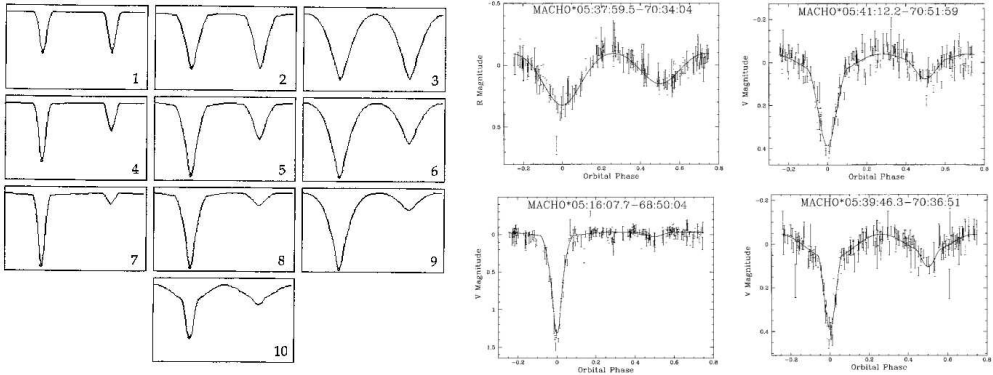} \\
\end{center}
\caption{\label{alcock} A decimal classification scheme proposed by \citeauthor{alcock1997}. The 
scheme relies of two physical parameters: the sum of relative radii and the surface brightness ratio. 
Four plots on the right are classified data from the MACHO survey. Adopted from \cite{alcock1997}.}
\end{figure}

Attacks on LMC continued by \cite{alcock1997}, who used the MACHO database \citep{macho1995} to 
analyse 611 bright EBs. The selection was based on visual identification by examining phase plots. 
They pointed out two physical quantities that, besides inclination, account for most variance in 
light curves: the sum of relative radii and the surface brightness ratio. For preliminary analysis 
the authors used the \citet{nelson1972} code and, following the GCVS designation types, they proposed 
a new decimal classification scheme depicted in Figure~\ref{alcock}.

\begin{figure}
\begin{center}
\includegraphics[width=0.49\textwidth]{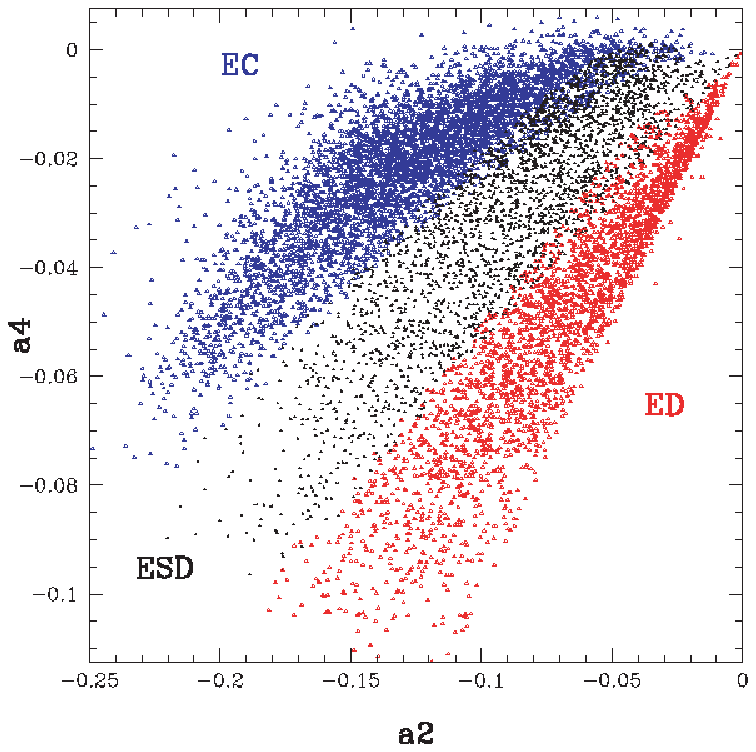}
\includegraphics[width=0.49\textwidth]{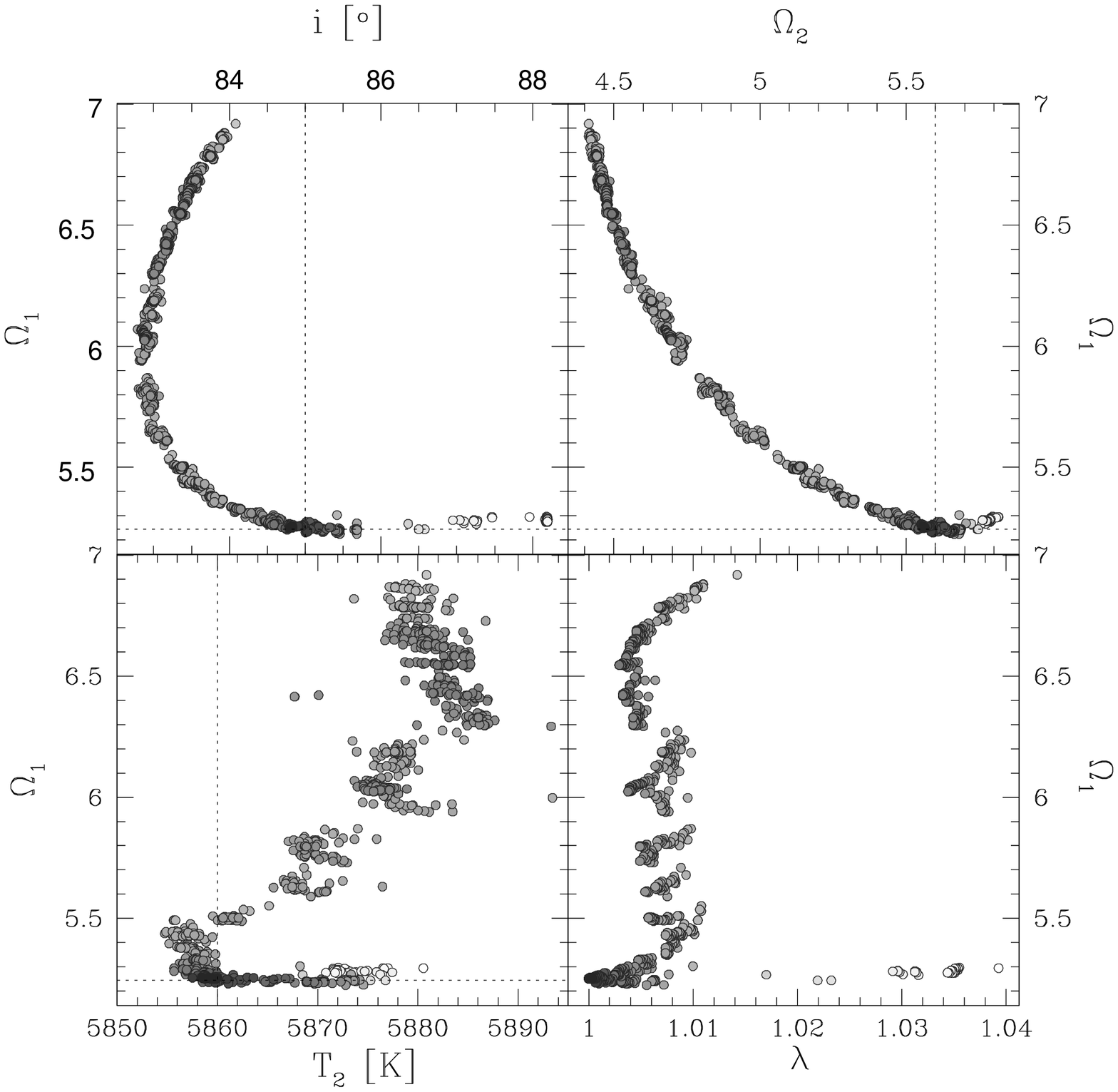} \\
\end{center}
\caption{\label{paczynski} Left: three types of eclipsing binaries (detached, semi-detached, and 
contact) mapped in the $a_2$--$a_4$ Fourier composition space; adopted from \citet{paczynski2006}. 
Right: heuristic scanning with Powell's direction set method. Converged results are shown for 
different parameter cross-sections; cross-hairs denote the right solution, and the symbol's shade of 
gray corresponds to the reached $\chi^2$ value: the darker the tone, the lower the $\chi^2$. Taken 
from \citet{prsa2006}.}
\end{figure}

The next survey to provide results for 933 EBs was OGLE \citep{szymanski1996}. Series of systematic 
analyses were conducted by \citet{rucinski1997a, rucinski1997b, rucinski1998} and later 
\cite{maceroni1999,rucinski2001} that stressed the success and importance of the Fourier decomposition 
technique (FDT) for classification of variable stars. The technique itself -- fitting a 4$^{\rm th}$ order 
Fourier series to phased data curves and mapping different types of variables in Fourier coefficient 
space (c.f., Figure~\ref{paczynski}, left) -- was first proposed for EBs already by \citet{rucinski1973} 
and has been used ever since, most notably for classifying ASAS data \citep{pojmanski2002, paczynski2006}.

Somewhat ironically, the first one to implement a fully automatic analysis pipeline for obtaining 
physical parameters of EBs was the most vocal advocate against any automated approaches: R.E.~Wilson. 
In their two papers, \citet{wyithe2001, wyithe2002} carried out an automatic search from 1459 EBs in 
the SMC detected by OGLE to find ideal distance estimators. WD was run in an automatic mode for the 
first time, although on a stripped level of complexity: the model assumed canonical values for 
physical parameters poorly defined by a single-passband photometric data: mass ratio $q=1$, argument 
of periastron $\omega = 0$ or $\pi$, the temperature of the secondary $T_2 = 15\,000$K, no spots, 
simple reflection, synchronous rotation etc. Yet for the first time, an automatic, decision-making 
pipeline was tested against synthetic data and then applied to observations. Despite several 
deficiencies (systematics introduced through assertions, DC-based method without heuristical search 
for solution uniqueness, no account of reddening) the authors succeeded to come up with two groups 
of candidates for ideal distance indicators: widely detached EBs and EBs with total eclipses. A 
manual follow-up analysis of 19 bright, large-amplitude candidates in their list was done e.g., by 
\citet{graczyk2003}, deriving the distance modulus to the \object{SMC} to be $\sim$18.9$\pm$0.1.

Meanwhile, a number of reliable solutions of individual EB solutions was steadily growing by a 
dedicated series of manual analyses, e.g., by \citeauthor{andersen1983}, \citeauthor{munari2001} and 
others. Instead of immediately going for survey data, our group decided to test fully automatic 
pipelines on these high-quality data. In our early work \citep{prsa2003} we obtained encouraging 
results for 5 morphologically different EBs, stressing importance of data diversity -- photometric 
data without RVs does not suffice for accurate modeling results. Trying to follow up on our devised 
scheme, we soon identified main deficiencies of the DC algorithm: since it is based on numerical 
derivatives, it may frequently diverge, and it gets stuck in local minima. To overcome this, we 
proposed two types of derivative-less methods: Nelder \& Mead's downhill Simplex method 
\citep{prsa2005b} and Powell's Direction set method \citep{prsa2006}. To understand and explore 
parameter degeneracy, heuristic scanning and parameter kicking were introduced (\citealt{phoebe2005}, 
c.f., Figure~\ref{paczynski}, right) -- the problem does \emph{not} lie in the DC, but in the inverse 
problem itself: its non-linearity, parameter degeneracy and data quality limitations. With this in 
mind we created a new modeling environment called PHOEBE\footnote{More information on PHOEBE may be 
found at {\tt http://phoebe.fiz.uni-lj.si}.} (PHysics Of Eclipsing BinariEs; \citealt{phoebe2005}) 
that features a flexible scripting language. This language is developed specifically with modeling 
and analysis of large surveys in mind.

\begin{figure}
\begin{center}
\includegraphics[width=0.561\textwidth]{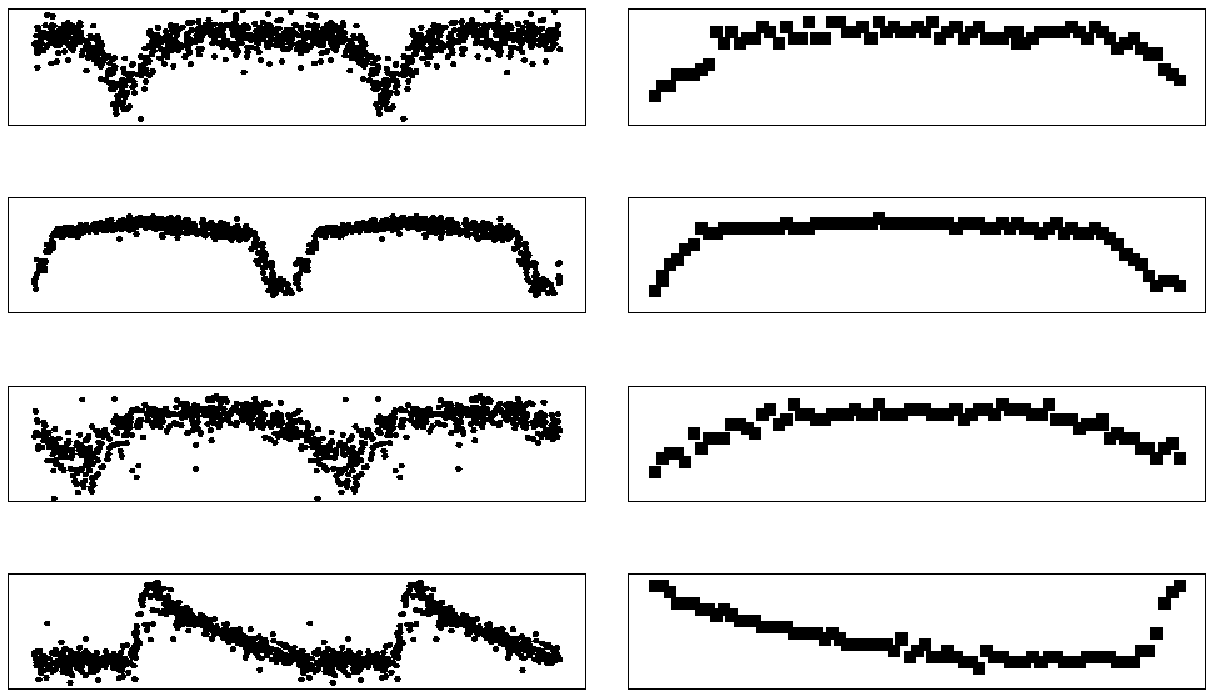}
\includegraphics[width=0.419\textwidth]{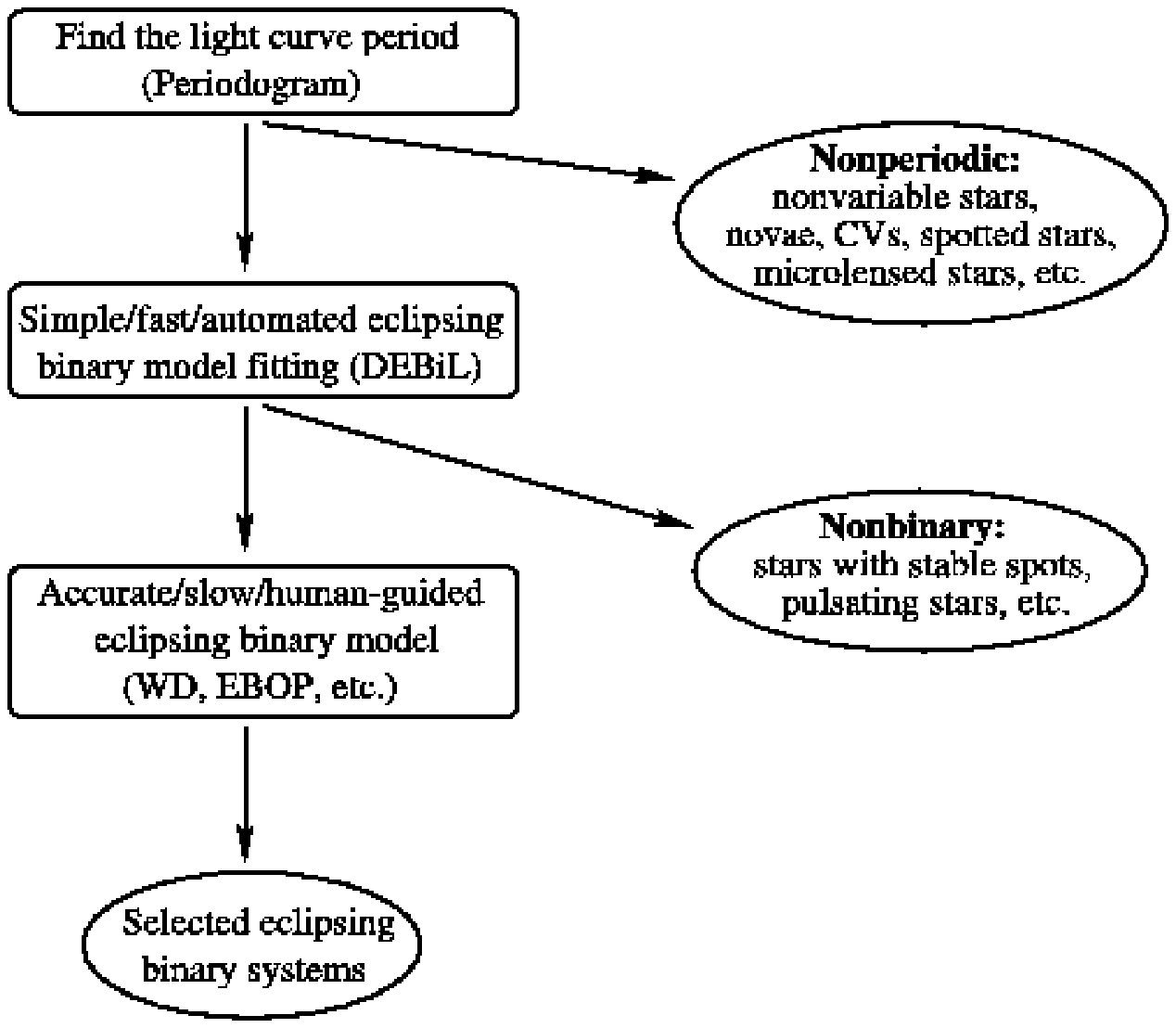}
\end{center}
\caption{\label{wyrzykowski} Left: an example of conversion of phased light curves to 70$\times$15 
pixel images, which are fed to the neural network image recognition algorithm. Taken from 
\citet{wyrzykowski2003b}. Right: a tier-based pipeline proposed by \citet{devor2005}: observed light 
curves are passed sequentially through filters and only the ones that fulfil all criteria make it to 
the next tier.}
\end{figure}

Continuing with the OGLE data harvest, \citet{wyrzykowski2003b, wyrzykowski2004} identified 2580 EBs 
in the LMC and 1351 EBs in the SMC. The novelty of their classification approach is using Artificial 
neural networks (ANN) as an image recognition algorithm, based on phased data curves that have been 
converted to low-resolution images as depicted on Figure~\ref{wyrzykowski}. Their classification 
pipeline was backed up by visual examinations of results. Although there were no physical analyses 
in their pipeline, observational properties of the sample, as well as 36 distance estimator 
candidates for the LMC, have been derived.

In 2005, \citeauthor{devor2005} implemented a tier-based elimination pipeline: observed light curves 
are sequentially passed through filters in the order of increasing computational time cost. Each 
tier filters out light curves that do not conform to the given criteria. Once a clean sample of light 
curves is available, it is submitted to a central part of the pipeline, a dedicated program DEBiL 
(Detached eclipsing binary light curve solver; c.f., Figure~\ref{wyrzykowski}, right), fitting a 
simplified EB model (spherical, limb-darkened stars on a classical Keplerian orbit) to observations. 
The pipeline yielded 10\,861 eclipsing binaries out of 218\,699 bulge field variables from OGLE II 
data \citep{udalski1997}. Its main advantage is speed: $\sim$ 1 minute per light curve on a 333MHz 
Sun UltraSparc 5 workstation. Its main deficiencies are lack of the 3rd proposed tier (accurate 
analysis) and an overly-simplified model that may produce false positives among other variables.

\begin{figure}
\begin{center}
\includegraphics[width=\textwidth]{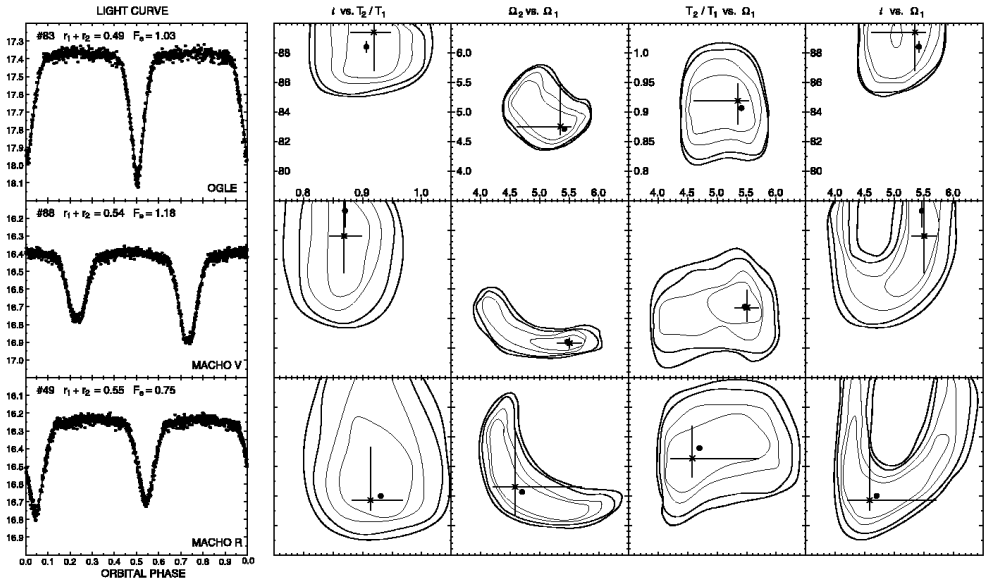}
\end{center}
\caption{\label{michalska} Examples of Monte-Carlo simulations for three EB light curves. Most 
importantly, the authors stress the importance of examining phase space cross-sections depicted on 
the right. Please refer to the original paper for further details. Taken from \citet{michalska2005}.}
\end{figure}

One of the best papers, in our opinion, that dealt with eclipsing binaries from OGLE II data, was the 
one by \citet{michalska2005}. Its thorough analysis and deep insight into caveats of the EB field 
make it exemplary for all similar undertakings in the future. The authors limited their analysis to 
bright ($V < 17.5$, $V-I < 0.5$), high S/N, EA type binaries that exhibit small proximity effects. 
After proving by example that the original differential image analysis (DIA) calibration is flawed 
due to uncertainty of reference flux in the flux-to-magnitude calibration, they proposed a novel 
method of calibrating DIA data and demonstrated its significantly better results. Once the OGLE II 
data has been re-calibrated, the authors added MACHO, OGLE I and EROS data (when available). The data 
have been submitted to a WD-based pipeline: the first step was to find initial parameter estimates by 
the Monte-Carlo method (c.f., Figure~\ref{michalska}), and the second step was to converge to the final 
solution by DC. A result is a list of 98 proposed candidates for distance estimates to the LMC, along 
with accurately determined parameters in relative units. Out of the sample, 58 stars are found to 
have eccentric orbits, and 14 systems are exhibiting apsidal motion.

Out of the crowd emerges yet another program to tackle the problem: EBAI (Eclipsing Binaries with 
Artificial Intelligence; \citealt{devinney2005}). This project does not only classify the data, it 
does more: blindingly fast, it determines coarse parameters of eclipsing binaries in a large data set. 
Study is underway for these parameters to be fed to a WD-based solver within PHOEBE. This solver maps 
the hyperspace around the solution, verifying its uniqueness and heuristically determining error 
estimates.

Another recent work that we wish to draw specific attention to has been done by \cite{tamuz2006}. The 
authors devised a new algorithm called EBAS (Eclipsing Binary Automatic Solver), aimed specifically 
to large datasets and thus based on the faster, yet less accurate EBOP code \citep{etzel1981}. 
Similarly to the discussed predecessors, EBAS also uses the sum of relative radii as a principal 
parameter. Yet there are two important novelties of their approach: instead of inclination the 
authors introduced the impact parameter -- the projected distance between the centers of the two 
stars during the primary eclipse, measured in terms of the sum of radii -- and they introduced a 
new ``alarm" statistics, the goal of which is to automatically discriminate best-fit $\chi^2$ values 
from still apparently acceptable values, but corresponding to distinctively wrong solutions. A 
follow-up application of EBAS on 938 OGLE LMC binaries with B-type main-sequence primary stars 
\citep{mazeh2006} yielded the distributions of the fractional radii of the two components and their 
sum, the brightness ratios and the periods of the short-period binaries. Intriguingly, they observed 
that the distribution in $\log P$ is \emph{flat} on the 2-10 days interval and that the detected 
frequency of their target stars is significantly smaller than the frequency deduced by dedicated RV 
surveys. The details on these findings are also given by \citet{mazeh2006b}.

Our attempt to preserve paper readability, and struggling against page limits at the same time, 
regrettably prohibits us to summarize all the work done so far. That is why we wish to at least 
acknowledge other important developments of this field -- and to apologize for any unintentional 
omissions in this brief review. Reader interested in pipeline reduction of binary light curves from 
large-scale surveys will surely benefit from the work of \citet{lastennet2002}, \citet{brett2004}, 
\citet{ribas2004}, \citet{wilson2004}, \citet{hilditch2004, hilditch2005}, \citet{eyer2005}, 
\citet{groenewegen2005}, \citet{naficy2005}, \citet{sarro2006} and many others.

\section{Traditional observations are \emph{not} obsolete}

After so much stress on surveys, missions and sophistication in fully automatic approaches it is 
tempting to conclude that traditional object-by-object observations have become obsolete. This is 
one of most dangerous misconceptions, apparently powered even by our own statement in the 
introduction that most (if not all) of the candidates have already been observed a number of times. 
Although these hot topics are appealing because of shear numbers of observed objects, there are 
several deficiencies in the context of eclipsing binaries that we should be aware of:

\begin{itemize}

\item Surveys and missions have a limited life-time that is generally not governed by the eclipsing 
binary harvest. Rather, limitations arise on account of funding, technology and reaching primary 
scientific objectives. A direct consequence is the selection effect in observed EBs: only the ones 
with suitable periods will have been detected.

\item The main driving idea of surveys is to acquire as much data as quickly as possible. Due to 
adopted sky scanning laws, the sky coverage is typically non-uniform and the observations are thus 
clustered in time. Although this might not seem too important for close binaries, it is critical in 
case of well detached binaries where there is practically no surface deformation and where eclipses 
occur only on a narrow phase interval. Having a point or two within the eclipse is hardly any 
different than having no point at all.

\item In order to reach survey completeness in terms of object counts during the mission life-time, 
the number of data points per object is usually poor. This means that the phase coverage for 
eclipsing binaries is often not sufficient for recognition and classification purposes, because of 
the strong sensitivity of period detection algorithms to phase completeness.

\item Unprecedented in numbers does not mean unprecedented in accuracy. At least so far, survey data 
has been more challenging to reduce and analyse than a typical dedicated observation because of the 
significant data scatter. However, WIRE (the Wide-field Infra-Red Explorer, \citealt{hacking1999}), 
despite its failure to achieve primary scientific goals, had already given us a snapshot of the 
milli-magnitude photometric precision of the future (c.f., Figure~\ref{wire}).

\begin{figure}
\begin{center}
\includegraphics[width=\textwidth]{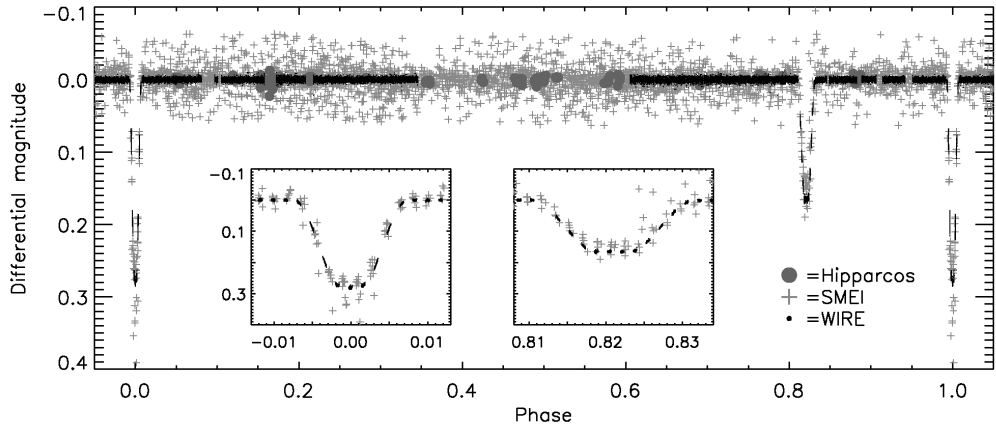} \\
\end{center}
\caption{\label{wire} Phased light curve of $\psi$ Cen. Data points from WIRE (black dots) may be 
compared to those from SMEI (gray plus symbols) and from Hipparcos (grey circles). Taken from 
\citet{bruntt2006}.}
\end{figure}

\item Most importantly: surveys usually lack data diversity. In order to get absolute temperatures 
of both stars and interstellar extinction, multi-passband photometry is needed; to get reliable 
estimates of absolute sizes of an eclipsing binary system, radial velocities are needed. To break 
inter--parameter correlations and solution degeneracy, as many diverse data-sets as possible are 
needed: astrometry and parallaxes, photometry, polarimetry, spectroscopy --- the more the better. 
Different physical and geometrical parameters, and their inter-dependencies, are revealed by 
different types of data-sets.

\end{itemize}

\noindent
If we take all of the above into account, we may only conclude that follow-up observations are still 
badly needed.

\section{Conclusions and prospects for the future}

Overwhelming data quantities are upon us and changing traditional ways of modeling and analysis of 
eclipsing binaries is thus inevitable. There are many fine studies that bring us closer to this goal. 
One of the greatest properties of astronomy, when compared to other sciences, is a strong sense of 
collaboration, and absence of blind competition, between astronomers; our questions, therefore, on 
how to facilitate and how to propagate the idea of joint development of these new approaches, and 
how to handle huge data-sets that are pouring in, are aimed at every single individual interested in 
contributing its own piece to this fascinating puzzle. These are indeed scientifically challenging 
times and it would be too bad if we missed them.


\begin{thebibliography}{63}
\expandafter\ifx\csname natexlab\endcsname\relax\def\natexlab#1{#1}\fi

\bibitem[{{Alcock} {\etal}(1997){Alcock}, {Allsman}, {Alves}, {Axelrod},
  {Becker}, {Bennett}, {Cook}, {Freeman}, {Griest}, {Lacy}, {Lehner},
  {Marshall}, {Minniti}, {Peterson}, {Pratt}, {Quinn}, {Rodgers}, {Stubbs},
  {Sutherland}, \& {Welch}}]{alcock1997}
{Alcock}, C., {Allsman}, R.A., {Alves}, D., {Axelrod}, T.S., {Becker}, A.C.,
  {Bennett}, D.P., {Cook}, K.H., {Freeman}, K.C., {Griest}, K., {Lacy},
  C.H.S., {Lehner}, M.J., {Marshall}, S.L., {Minniti}, D., {Peterson},
  B.A., {Pratt}, M.R., {Quinn}, P.J., {Rodgers}, A.W., {Stubbs}, C.W.,
  {Sutherland}, W., \& {Welch}, D.L. 1997, \textit{AJ}, 114, 326

\bibitem[{{Andersen} {\etal}(1983){Andersen}, {Clausen}, {Nordstroem}, \&
  {Reipurth}}]{andersen1983}
{Andersen}, J., {Clausen}, J.V., {Nordstroem}, B., \& {Reipurth}, B. 1983,
  \textit{A\&A}, 121, 271

\bibitem[{{Baglin} {\etal}(2002){Baglin}, {Auvergne}, {Barge}, {Buey},
  {Catala}, {Michel}, {Weiss}, \& {COROT Team}}]{corot2002}
{Baglin}, A., {Auvergne}, M., {Barge}, P., {Buey}, J.-T., {Catala}, C.,
  {Michel}, E., {Weiss}, W., \& {COROT Team}. 2002, in \textit{Stellar
  Structure and Habitable Planet Finding}, ESA SP-485, ed. B.~{Battrick}, F.~{Favata}, 
  I.W.~{Roxburgh}, \& D.~{Galadi}, 17--24

\bibitem[{{Brett} {\etal}(2004){Brett}, {West}, \& {Wheatley}}]{brett2004}
{Brett}, D.R., {West}, R.G., \& {Wheatley}, P.J. 2004, \textit{MNRAS}, 353, 369

\bibitem[{{Bruntt} {\etal}(2006){Bruntt}, {Southworth}, {Torres}, {Penny},
  {Clausen}, \& {Buzasi}}]{bruntt2006}
{Bruntt}, H., {Southworth}, J., {Torres}, G., {Penny}, A.J., {Clausen}, J.V.,
  \& {Buzasi}, D.L. 2006, \textit{A\&A}, 456, 651

\bibitem[{{Cook} {\etal}(1995){Cook}, {Alcock}, {Allsman}, {Axelrod},
  {Freeman}, {Peterson}, {Quinn}, {Rodgers}, {Bennett}, {Reimann}, {Griest},
  {Marshall}, {Pratt}, {Stubbs}, {Sutherland}, \& {Welch}}]{macho1995}
{Cook}, K.H., {Alcock}, C., {Allsman}, H.A., {Axelrod}, T.S., {Freeman},
  K.C., {Peterson}, B.A., {Quinn}, P.J., {Rodgers}, A.W., {Bennett}, D.P.,
  {Reimann}, J., {Griest}, K., {Marshall}, S.L., {Pratt}, M.R., {Stubbs},
  C.W., {Sutherland}, W., \& {Welch}, D.L. 1995, in IAU
  Colloq. 155, \textit{Astrophysical Applications of Stellar Pulsation}, ASP Conf. Ser. 83, ed. R.
  {Stobie} \& P. {Whitelock}, 221

\bibitem[{{Devinney} {\etal}(2005){Devinney}, {Guinan}, {Bradstreet},
  {DeGeorge}, {Giammarco}, {Alcock}, \& {Engle}}]{devinney2005}
{Devinney}, E., {Guinan}, E., {Bradstreet}, D., {DeGeorge}, M., {Giammarco},
  J., {Alcock}, C., \& {Engle}, S. 2005, \textit{BAAS}, 1212

\bibitem[{{Devor}(2005)}]{devor2005}
{Devor}, J. 2005, \textit{ApJ}, 628, 411

\bibitem[{{Droege} {\etal}(2006){Droege}, {Richmond}, {Sallman}, \&
  {Creager}}]{droege2006}
{Droege}, T.F., {Richmond}, M.W., {Sallman}, M.P., \& {Creager}, R.P. 2006,
  \textit{ArXiv Astrophysics} e-prints

\bibitem[{{Etzel}(1981)}]{etzel1981}
{Etzel}, P.B. 1981, in \textit{Photometric and Spectroscopic Binary Systems}, ed. E.B.
  {Carling} \& Z.~{Kopal}, 111

\bibitem[{{Eyer} \& {Blake}(2005)}]{eyer2005}
{Eyer}, L. \& {Blake}, C. 2005, \textit{MNRAS}, 358, 30

\bibitem[{{Friedemann} {\etal}(1996){Friedemann}, {Guertler}, \&
  {Loewe}}]{friedemann1996}
{Friedemann}, C., {Guertler}, J., \& {Loewe}, M. 1996, \textit{A\&AS}, 117, 205

\bibitem[{{Graczyk}(2003)}]{graczyk2003}
{Graczyk}, D. 2003, \textit{MNRAS}, 342, 1334

\bibitem[{{Grison} {\etal}(1995){Grison}, {Beaulieu}, {Pritchard}, {Tobin},
  {Ferlet}, {Vidal-Madjar}, {Guibert}, {Alard}, {Moreau}, {Tajahmady},
  {Maurice}, {Prevot}, {Gry}, {Aubourg}, {Bareyre}, {Brehin}, {Gros},
  {Lachieze-Rey}, {Laurent}, {Lesquoy}, {Magneville}, {Milsztajn}, {Moscoso},
  {Queinnec}, {Renault}, {Rich}, {Spiro}, {Vigroux}, {Zylberajch}, {Ansari},
  {Cavalier}, \& {Moniez}}]{grison1995}
{Grison}, P., {Beaulieu}, J.-P., {Pritchard}, J.D., {Tobin}, W., {Ferlet}, R.,
  {Vidal-Madjar}, A., {Guibert}, J., {Alard}, C., {Moreau}, O., {Tajahmady},
  F., {Maurice}, E., {Prevot}, L., {Gry}, C., {Aubourg}, E., {Bareyre}, P.,
  {Brehin}, S., {Gros}, M., {Lachieze-Rey}, M., {Laurent}, B., {Lesquoy}, E.,
  {Magneville}, C., {Milsztajn}, A., {Moscoso}, L., {Queinnec}, F., {Renault},
  C., {Rich}, J., {Spiro}, M., {Vigroux}, L., {Zylberajch}, S., {Ansari}, R.,
  {Cavalier}, F., \& {Moniez}, M.1995, \textit{A\&AS}, 109, 447

\bibitem[{{Groenewegen}(2005)}]{groenewegen2005}
{Groenewegen}, M.A.T. 2005, \textit{A\&A}, 439, 559

\bibitem[{{Hacking} {\etal}(1999){Hacking}, {Lonsdale}, {Gautier}, {Herter},
  {Shupe}, {Stacey}, {Fang}, {Xu}, {Graf}, {Werner}, {Soifer}, {Moseley}, \&
  {Houck}}]{hacking1999}
{Hacking}, P., {Lonsdale}, C., {Gautier}, T., {Herter}, T., {Shupe}, D.,
  {Stacey}, G., {Fang}, F., {Xu}, C., {Graf}, P., {Werner}, M., {Soifer}, B.,
  {Moseley}, H., \& {Houck}, J. 1999, in \textit{Astrophysics with
  Infrared Surveys: A Prelude to SIRTF}, ASP Conf. Ser. 177, ed. M.D. {Bicay}, R.M. {Cutri}, \&
  B.F. {Madore}, 409

\bibitem[{{Hadrava}(1990)}]{hadrava1990}
{Hadrava}, P. 1990, \textit{Contributions of the Astronomical Observatory Skalnate
  Pleso}, 20, 23

\bibitem[{{Hilditch} {\etal}(2004){Hilditch}, {Harries}, \&
  {Howarth}}]{hilditch2004}
{Hilditch}, R.W., {Harries}, T.J., \& {Howarth}, I.D. 2004, \textit{New Astronomy
  Review}, 48, 687

\bibitem[{{Hilditch} {\etal}(2005){Hilditch}, {Howarth}, \&
  {Harries}}]{hilditch2005}
{Hilditch}, R.W., {Howarth}, I.D., \& {Harries}, T.J. 2005, \textit{MNRAS}, 357, 304

\bibitem[{{Hill}(1979)}]{hill1979}
{Hill}, G. 1979, \textit{Publications of the Dominion Astrophysical Observatory
  Victoria}, 15, 297

\bibitem[{{Hor{\'a}k}(1966)}]{horak1966}
{Hor{\'a}k}, T. 1966, \textit{Bulletin of the Astronomical Institutes of
  Czechoslovakia}, 17, 27

\bibitem[{{Hor{\'a}k}(1970)}]{horak1970}
{Hor{\'a}k}, T.B. 1970, \textit{AJ}, 75, 1116

\bibitem[{{Kaiser} {\etal}(2002){Kaiser}, {Aussel}, {Burke}, {Boesgaard},
  {Chambers}, {Chun}, {Heasley}, {Hodapp}, {Hunt}, {Jedicke}, {Jewitt},
  {Kudritzki}, {Luppino}, {Maberry}, {Magnier}, {Monet}, {Onaka}, {Pickles},
  {Rhoads}, {Simon}, {Szalay}, {Szapudi}, {Tholen}, {Tonry}, {Waterson}, \&
  {Wick}}]{panstarrs2002}
{Kaiser}, N., {Aussel}, H., {Burke}, B.E., {Boesgaard}, H., {Chambers}, K.,
  {Chun}, M.R., {Heasley}, J.N., {Hodapp}, K.-W., {Hunt}, B., {Jedicke}, R.,
  {Jewitt}, D., {Kudritzki}, R., {Luppino}, G.A., {Maberry}, M., {Magnier},
  E., {Monet}, D.G., {Onaka}, P.M., {Pickles}, A.J., {Rhoads}, P.H.H.,
  {Simon}, T., {Szalay}, A., {Szapudi}, I., {Tholen}, D.J., {Tonry}, J.L.,
  {Waterson}, M., \& {Wick}, J. 2002, in \textit{Survey and Other Telescope
  Technologies and Discoveries}, Proceedings of the SPIE, Volume 4836, ed. J.A. {Tyson}
  \& S.~{Wolff}, 154--164

\bibitem[{{Kholopov} {\etal}(1992){Kholopov}, {Samus}, {Durlevich},
  {Kazarovets}, {Kireeva}, \& {Tsvetkova}}]{gcvs1992}
{Kholopov}, P.N., {Samus}, N.N., {Durlevich}, O.V., {Kazarovets}, E.V.,
  {Kireeva}, N.N., \& {Tsvetkova}, T.M. 1992, \textit{Bulletin d'Information du
  Centre de Donnees Stellaires}, 40, 15

\bibitem[{{Koch} {\etal}(2004){Koch}, {Borucki}, {Dunham}, {Geary},
  {Gilliland}, {Jenkins}, {Latham}, {Bachtell}, {Berry}, {Deininger}, {Duren},
  {Gautier}, {Gillis}, {Mayer}, {Miller}, {Shafer}, {Sobeck}, {Stewart}, \&
  {Weiss}}]{koch2004}
{Koch}, D.G., {Borucki}, W., {Dunham}, E., {Geary}, J., {Gilliland}, R.,
  {Jenkins}, J., {Latham}, D., {Bachtell}, E., {Berry}, D., {Deininger}, W.,
  {Duren}, R., {Gautier}, T.N., {Gillis}, L., {Mayer}, D., {Miller}, C.D.,
  {Shafer}, D., {Sobeck}, C.K., {Stewart}, C., \& {Weiss}, M. 2004, in
  \textit{Optical, Infrared, and Millimeter Space Telescopes}. 
  Proceedings of the SPIE, Volume 5487, ed. J.C.
  {Mather}, 1491--1500

\bibitem[{{Lammers}(2005)}]{lammers2005}
{Lammers}, U. 2005, in ESA SP-576: \textit{The Three-Dimensional Universe with Gaia},
  ed. C.~{Turon}, K.S. {O'Flaherty}, \& M.A.C. {Perryman}, 445

\bibitem[{{Lastennet} \& {Valls-Gabaud}(2002)}]{lastennet2002}
{Lastennet}, E. \& {Valls-Gabaud}, D. 2002, \textit{A\&A}, 396, 551

\bibitem[{{Linnell} \& {Hubeny}(1994)}]{linnell1994}
{Linnell}, A.P. \& {Hubeny}, I. 1994, \textit{ApJ}, 434, 738

\bibitem[{{Maceroni} \& {Rucinski}(1999)}]{maceroni1999}
{Maceroni}, C., \& {Rucinski}, S.M. 1999, \textit{AJ}, 118, 1819

\bibitem[{{Mazeh} {\etal}(2006{\natexlab{a}}){Mazeh}, {Tamuz}, \&
  {North}}]{mazeh2006}
{Mazeh}, T., {Tamuz}, O., \& {North}, P. 2006{\natexlab{a}}, \textit{MNRAS}, 367, 1531

\bibitem[{{Mazeh} {\etal}(2006{\natexlab{b}}){Mazeh}, {Tamuz}, \&
  {North}}]{mazeh2006b}
---. 2006{\natexlab{b}}, \textit{Ap\&SS}, 44

\bibitem[{{Michalska} \& {Pigulski}(2005)}]{michalska2005}
{Michalska}, G. \& {Pigulski}, A. 2005, \textit{A\&A}, 434, 89

\bibitem[{{Mochnacki} \& {Doughty}(1972)}]{mochnacki1972}
{Mochnacki}, S.W. \& {Doughty}, N.A. 1972, \textit{MNRAS}, 156, 51

\bibitem[{{Munari} {\etal}(2001){Munari}, {Tomov}, {Zwitter}, {Milone},
  {Kallrath}, {Marrese}, {Boschi}, {Pr{\v s}a}, {Tomasella}, \&
  {Moro}}]{munari2001}
{Munari}, U., {Tomov}, T., {Zwitter}, T., {Milone}, E.F., {Kallrath}, J.,
  {Marrese}, P.M., {Boschi}, F., {Pr{\v s}a}, A., {Tomasella}, L., \& {Moro},
  D. 2001, \textit{A\&A}, 378, 477

\bibitem[{{Naficy} {\etal}(2005){Naficy}, {Riazi}, \&
  {Kiasatpour}}]{naficy2005}
{Naficy}, K., {Riazi}, N., \& {Kiasatpour}, A. 2005, \textit{AJ}, 130, 1862

\bibitem[{{Nelson} \& {Davis}(1972)}]{nelson1972}
{Nelson}, B. \& {Davis}, W.D. 1972, \textit{ApJ}, 174, 617

\bibitem[{{Neugebauer} {\etal}(1984){Neugebauer}, {Habing}, {van Duinen},
  {Aumann}, {Baud}, {Beichman}, {Beintema}, {Boggess}, {Clegg}, {de Jong},
  {Emerson}, {Gautier}, {Gillett}, {Harris}, {Hauser}, {Houck}, {Jennings},
  {Low}, {Marsden}, {Miley}, {Olnon}, {Pottasch}, {Raimond}, {Rowan-Robinson},
  {Soifer}, {Walker}, {Wesselius}, \& {Young}}]{iras1984}
{Neugebauer}, G., {Habing}, H.J., {van Duinen}, R., {Aumann}, H.H., {Baud},
  B., {Beichman}, C.A., {Beintema}, D.A., {Boggess}, N., {Clegg}, P.E., {de
  Jong}, T., {Emerson}, J.P., {Gautier}, T.N., {Gillett}, F.C., {Harris},
  S., {Hauser}, M.G., {Houck}, J.R., {Jennings}, R.E., {Low}, F.J.,
  {Marsden}, P.L., {Miley}, G., {Olnon}, F.M., {Pottasch}, S.R., {Raimond},
  E., {Rowan-Robinson}, M., {Soifer}, B.T., {Walker}, R.G., {Wesselius},
  P.R., \& {Young}, E. 1984, \textit{ApJL}, 278, L1

\bibitem[{{Paczy{\'n}ski} {\etal}(2006){Paczy{\'n}ski}, {Szczygie{\l}},
  {Pilecki}, \& {Pojma{\'n}ski}}]{paczynski2006}
{Paczy{\'n}ski}, B., {Szczygie{\l}}, D.M., {Pilecki}, B., \& {Pojma{\'n}ski},
  G. 2006, \textit{MNRAS}, 368, 1311

\bibitem[{{Palanque-Delabrouille} {\etal}(1998){Palanque-Delabrouille},
  {Afonso}, {Albert}, {Andersen}, {Ansari}, {Aubourg}, {Bareyre}, {Bauer},
  {Beaulieu}, {Bouquet}, {Char}, {Charlot}, {Couchot}, {Coutures}, {Derue},
  {Ferlet}, {Glicenstein}, {Goldman}, {Gould}, {Graff}, {Gros}, {Haissinski},
  {Hamilton}, {Hardin}, {de Kat}, {Lesquoy}, {Loup}, {Magneville}, {Mansoux},
  {Marquette}, {Maurice}, {Milsztajn}, {Moniez}, {Perdereau}, {Prevot},
  {Renault}, {Rich}, {Spiro}, {Vidal-Madjar}, {Vigroux}, {Zylberajch}, \& {The
  EROS Collaboration}}]{pd1998}
{Palanque-Delabrouille}, N., {Afonso}, C., {Albert}, J.N., {Andersen}, J.,
  {Ansari}, R., {Aubourg}, E., {Bareyre}, P., {Bauer}, F. \etal\ \& {the EROS
  Collaboration}. 1998, \textit{A\&A}, 332, 1

\bibitem[{{Payne-Gaposchkin}(1971)}]{payne1971}
{Payne-Gaposchkin}, C.H. 1971, \textit{The variable stars of the Large Magellanic
  Cloud}, (Smithsonian Contributions to Astrophysics, Washington: Smithsonian
  Institution Press, |c1971)

\bibitem[{{Perryman} {\etal}(2001){Perryman}, {de Boer}, {Gilmore}, {H{\o}g},
  {Lattanzi}, {Lindegren}, {Luri}, {Mignard}, {Pace}, \& {de
  Zeeuw}}]{perryman2001}
{Perryman}, M.A.C., {de Boer}, K.S., {Gilmore}, G., {H{\o}g}, E.,
  {Lattanzi}, M.G., {Lindegren}, L., {Luri}, X., {Mignard}, F., {Pace}, O., \&
  {de Zeeuw}, P.T. 2001, \textit{A\&A}, 369, 339

\bibitem[{{Perryman} \& {ESA}(1997)}]{hipparcos1997}
{Perryman}, M.A.C. \& {ESA}. 1997, \textit{The Hipparcos and Tycho catalogues.~Astrometric
  and photometric star catalogues derived from the ESA Hipparcos Space
  Astrometry Mission}, Publisher: Noordwijk, Netherlands: ESA Publications
  Division, Series: ESA SP Series vol no: 1200, ISBN: 9290923997 (set)

\bibitem[{{Pojmanski}(1997)}]{pojmanski1997}
{Pojmanski}, G. 1997, \textit{Acta Astronomica}, 47, 467

\bibitem[{{Pojmanski}(2002)}]{pojmanski2002}
---. 2002, \textit{Acta Astronomica}, 52, 397

\bibitem[{{Pr{\v s}a}(2003)}]{prsa2003}
{Pr{\v s}a}, A. 2003, in \textit{GAIA Spectroscopy: Science and
  Technology}, ASP Conf. Ser. 298, ed. U.~{Munari}, 457

\bibitem[{{Pr{\v s}a} \& {Zwitter}(2005{\natexlab{a}})}]{phoebe2005}
{Pr{\v s}a}, A. \& {Zwitter}, T. 2005{\natexlab{a}}, \textit{ApJ}, 628, 426

\bibitem[{{Pr{\v s}a} \& {Zwitter}(2005{\natexlab{b}})}]{prsa2005a}
---. 2005{\natexlab{b}}, \textit{Ap\&SS}, 296, 315

\bibitem[{{Pr{\v s}a} \& {Zwitter}(2005{\natexlab{c}})}]{prsa2005b}
{Pr{\v s}a}, A., \& {Zwitter}, T. 2005{\natexlab{c}}, in \textit{The
  Three-Dimensional Universe with Gaia}, ESA SP-576, ed. C.~{Turon}, K.S. {O'Flaherty}, \&
  M.A.C. {Perryman}, 611

\bibitem[{{Pr{\v s}a} \& {Zwitter}(2006)}]{prsa2006}
---. 2006, \textit{ArXiv Astrophysics} e-prints

\bibitem[{{Ribas} {\etal}(2004){Ribas}, {Jordi}, {Vilardell}, {Gim{\'e}nez},
  \& {Guinan}}]{ribas2004}
{Ribas}, I., {Jordi}, C., {Vilardell}, F., {Gim{\'e}nez}, {\'A}., \& {Guinan},
  E.F. 2004, \textit{New Astronomy Review}, 48, 755

\bibitem[{{Rucinski}(1973)}]{rucinski1973}
{Rucinski}, S.M. 1973, \textit{Acta Astronomica}, 23, 79

\bibitem[{{Rucinski}(1997{\natexlab{a}})}]{rucinski1997b}
{Rucinski}, S.M. 1997{\natexlab{a}}, \textit{AJ}, 113, 1112

\bibitem[{{Rucinski}(1997{\natexlab{b}})}]{rucinski1997a}
---. 1997{\natexlab{b}}, \textit{AJ}, 113, 407

\bibitem[{{Rucinski}(1998)}]{rucinski1998}
---. 1998, \textit{AJ}, 115, 1135

\bibitem[{{Rucinski} \& {Maceroni}(2001)}]{rucinski2001}
{Rucinski}, S.M. \& {Maceroni}, C. 2001, \textit{AJ}, 121, 254

\bibitem[{{Sarro} {\etal}(2006){Sarro}, {S{\'a}nchez-Fern{\'a}ndez}, \&
  {Gim{\'e}nez}}]{sarro2006}
{Sarro}, L.M., {S{\'a}nchez-Fern{\'a}ndez}, C., \& {Gim{\'e}nez}, {\'A}. 2006,
  \textit{A\&A}, 446, 395

\bibitem[{{Szymanski} {\etal}(1996){Szymanski}, {Udalski}, {Kubiak},
  {Kaluzny}, {Mateo}, \& {Krzeminski}}]{szymanski1996}
{Szymanski}, M., {Udalski}, A., {Kubiak}, M., {Kaluzny}, J., {Mateo}, M., \&
  {Krzeminski}, W. 1996, \textit{Acta Astronomica}, 46, 1

\bibitem[{{Tamuz} {\etal}(2006){Tamuz}, {Mazeh}, \& {North}}]{tamuz2006}
{Tamuz}, O., {Mazeh}, T., \& {North}, P. 2006, \textit{MNRAS}, 367, 1521

\bibitem[{{Tamuz} {\etal}(2005){Tamuz}, {Mazeh}, \& {Zucker}}]{tamuz2005}
{Tamuz}, O., {Mazeh}, T., \& {Zucker}, S. 2005, \textit{MNRAS}, 356, 1466

\bibitem[{{Tody}(1986)}]{tody1986}
{Tody}, D. 1986, in Proceedings of \textit{Instrumentation in astronomy VI}; 
  Part 2 (A87-36376 15-35). Bellingham,
  WA, Society of Photo-Optical Instrumentation Engineers, ed.
  D.L. {Crawford}, 733

\bibitem[{{Udalski} {\etal}(1997){Udalski}, {Kubiak}, \&
  {Szymanski}}]{udalski1997}
{Udalski}, A., {Kubiak}, M., \& {Szymanski}, M. 1997, \textit{Acta Astronomica}, 47, 319

\bibitem[{{Wilson}(2004)}]{wilson2004}
{Wilson}, R.E. 2004, \textit{New Astronomy Review}, 48, 695

\bibitem[{{Wilson} \& {Devinney}(1971)}]{wd1971}
{Wilson}, R.E. \& {Devinney}, E.J. 1971, \textit{ApJ}, 166, 605

\bibitem[{{Wood}(1971)}]{wood1971}
{Wood}, D.B. 1971, \textit{AJ}, 76, 701

\bibitem[{{Wyithe} \& {Wilson}(2001)}]{wyithe2001}
{Wyithe}, J.S.B. \& {Wilson}, R.E. 2001, \textit{ApJ}, 559, 260

\bibitem[{{Wyithe} \& {Wilson}(2002)}]{wyithe2002}
---. 2002, \textit{ApJ}, 571, 293

\bibitem[{{Wyrzykowski} {\etal}(2003){Wyrzykowski}, {Udalski}, {Kubiak},
  {Szymanski}, {Zebrun}, {Soszynski}, {Wozniak}, {Pietrzynski}, \&
  {Szewczyk}}]{wyrzykowski2003b}
{Wyrzykowski}, L., {Udalski}, A., {Kubiak}, M., {Szymanski}, M., {Zebrun}, K.,
  {Soszynski}, I., {Wozniak}, P.R., {Pietrzynski}, G., \& {Szewczyk}, O. 2003,
  \textit{Acta Astronomica}, 53, 1

\bibitem[{{Wyrzykowski} {etal.}(2004){Wyrzykowski}, {Udalski}, {Kubiak},
  {Szymanski}, {Zebrun}, {Soszynski}, {Wozniak}, {Pietrzynski}, \&
  {Szewczyk}}]{wyrzykowski2004}
{Wyrzykowski}, L., {Udalski}, A., {Kubiak}, M., {Szymanski}, M.K., {Zebrun},
  K., {Soszynski}, I., {Wozniak}, P.R., {Pietrzynski}, G., \& {Szewczyk}, O.
  2004, \textit{Acta Astronomica}, 54, 1

\end{thebibliography}
\end{document}